\documentclass{article}

\usepackage{dcase2021}
\usepackage{amsmath}
\usepackage{booktabs}
\usepackage{graphicx}
\usepackage{tabularx}
\usepackage{times}
\usepackage{url}

\usepackage[utf8]{inputenc}
\usepackage{csquotes}
\usepackage[T1]{fontenc}
\usepackage{mathtools}
\usepackage{microtype}
\usepackage{siunitx}

\graphicspath{{figures/}}

\newcommand{\dtags}{d_{\mathrm{tags}}}
\newcommand{\ddesc}{d_{\mathrm{desc}}}

\DeclareMathOperator*{\argmax}{arg\,max}
\DeclarePairedDelimiter{\norm}{\lVert}{\rVert}

\newcommand{\score}[2]{\((#1\)\small\(\pm #2\)\normalsize\()\)\%}

\title{ARCA23K: An audio dataset for investigating open-set label noise}

\name{Turab Iqbal, Yin Cao, Andrew Bailey, Mark D. Plumbley, Wenwu Wang}
\address{Centre for Vision, Speech and Signal Processing (CVSSP), University of Surrey, UK\\
         \{t.iqbal, yin.cao, andrew.bailey, m.plumbley, w.wang\}@surrey.ac.uk}

\begin{document}

\ninept
\maketitle

\begin{sloppy}

\begin{abstract}
The availability of audio data on sound sharing platforms such as Freesound
gives users access to large amounts of annotated audio. Utilising such data for
training is becoming increasingly popular, but the problem of label noise that
is often prevalent in such datasets requires further investigation. This paper
introduces ARCA23K, an Automatically Retrieved and Curated Audio dataset
comprised of over \num{23000} labelled Freesound clips. Unlike past datasets
such as FSDKaggle2018 and FSDnoisy18K, ARCA23K facilitates the study of label
noise in a more controlled manner. We describe the entire process of creating
the dataset such that it is fully reproducible, meaning researchers can extend
our work with little effort. We show that the majority of labelling errors in
ARCA23K are due to out-of-vocabulary audio clips, and we refer to this type of
label noise as open-set label noise. Experiments are carried out in which we
study the impact of label noise in terms of classification performance and
representation learning.

\end{abstract}


\section{Introduction}
\label{sec:intro}

Labelled audio data is a relatively scarce resource, yet it is vital for
training audio classifiers in a supervised fashion. With the emergence of
online sharing platforms such as Freesound \cite{freesound_fonseca} and YouTube
\cite{audioset_gemmeke}, users now have access to massive amounts of annotated
audio, and it is becoming increasingly popular to utilise this data for
training. For classifying general sound events, early examples of web-sourced
datasets include ESC-50 \cite{esc_piczak} and UrbanSound8K
\cite{urbansound_salamon}. However, these datasets are relatively small, which
is largely due to the high cost of manually verifying the data to ensure the
sounds are relevant and the labels are correct. At the time of writing, the
largest sound event dataset with thoroughly verified labels is FSD50K
\cite{fsd50k_fonseca}, which contains approximately \num{50000} sounds and is
the result of several years of crowdsourced labelling \cite{freesound_fonseca}.

Given the cost of label verification, there has been interest in reducing or
eliminating this aspect of dataset curation. AudioSet \cite{audioset_gemmeke},
for example, is a large-scale audio dataset comprised of over two million sounds
across hundreds of classes. AudioSet classes belong to an ontology in which the
classes share parent-child relationships. Although AudioSet clips have been
manually verified by listeners, the process was not thorough, and many labelling
errors remain \cite{noise_zhu}. The labels of other datasets, such as VGGSound
\cite{vggsound_chen}, have not been verified at all. In the case of
FSDKaggle2018 \cite{fsdkaggle2018_fonseca}, FSDKaggle2019
\cite{fsdkaggle2019_fonseca}, and FSDnoisy18k \cite{fsdnoisy18k_fonseca}, only
a small subset of the dataset has been manually verified. Nevertheless, these
datasets are attractive because they are relatively large. The challenge is
that the presence of labelling errors, or \textit{label noise}, can
significantly impact learning \cite{fsdnoisy18k_fonseca}. Hence, studying the
effects of label noise is important.

Due to label noise, rather than the training examples being drawn from the true
distribution, \(P\), examples are drawn from a corrupted distribution, \(Q\). In
the literature, the noise process responsible for this corruption is typically
assumed to be reversible, such that any incorrectly-labelled instance can be
relabelled \cite{noise_patrini, noise_sukhbaatar}. This is \textit{not} a
realistic assumption when retrieving and labelling web data, as the sounds that
are retrieved can be \textit{out-of-vocabulary (OOV)}
\cite{fsdnoisy18k_fonseca}. OOV sounds are sounds that do not belong to any of
the classes of interest. We refer to this type of label noise as
\textit{open-set label noise}. There is currently a disconnect where much of the
analysis and tools are for closed-set label noise, while open-set label noise
has received little attention in this respect. While there are works that
address datasets with open-set label noise \cite{ood_iqbal, noise_kumar,
noise_zhu, fsdkaggle2019_fonseca, fsdnoisy18k_fonseca}, the analysis is limited
by the lack of empirical insight.

In this paper, we introduce
\textit{ARCA23K}\footnote{\url{https://zenodo.org/record/5117901}}
(Automatically Retrieved and Curated Audio 23K), which is a dataset containing
more than \num{51} hours of audio data across \num{23727} Freesound clips and
\num{70} classes taken from the AudioSet ontology. The clips comprising the
training set have been retrieved and curated using an entirely automated
process, while the validation set and test set are subsets of FSD50K. Given the
absence of human verification, labelling errors are to be expected in the
training set. In particular, many of the audio clips are out-of-vocabulary.

Our aim in creating ARCA23K is to facilitate the study of real-world, open-set
label noise, including its effects on learning and how these effects can be
mitigated. Unlike datasets such as FSDnoisy18k, ARCA23K allows studying label
noise in a more controlled manner. Instead of manually verifying a subset of the
dataset, we introduce another dataset called \textit{ARCA23K-FSD}, which is a
subset of FSD50K. ARCA23K-FSD is essentially a `clean' counterpart of ARCA23K.
Under certain assumptions, this setup allows controlling the amount of label
noise by substituting clips from one dataset with clips from the other. A
similar idea was proposed in the image domain \cite{noise_jiang}.

The contributions of this paper are four-fold. First, we provide a detailed
description of how the ARCA23K datasets were created and release the
associated source code\footnote{\url{https://github.com/tqbl/arca23k-dataset}}.
Our intention is to provide a method of dataset creation that is realistic while
also being easily reproducible\footnote{Some clips on Freesound may be deleted,
which we cannot control.}, such that anyone can adopt or improve our method for
their own needs. Our second contribution is the release of ARCA23K itself (along with
ARCA23K-FSD). As all the clips are available under a Creative Commons license, we
are able to distribute the clips freely. Third, we characterise the label noise
present in ARCA23K by running listening tests. Finally, we conduct experiments
to examine the impact of open-set label noise on training audio classifiers,
which includes comparisons to synthetic label noise and an evaluation of the
representations that are learned.

\newpage

\section{ARCA23K-FSD}
\label{sec:arca23k-fsd}

In order to investigate open-set label noise, we propose two datasets: a clean
dataset with training examples drawn from \(P\) and a noisy dataset with
training examples from \(Q\). The number of examples per class is set to be
equal across the two datasets. By satisfying this requirement, we are able to
emulate the noise process that corrupts \(P\) to give \(Q\). More specifically,
a training example drawn from the clean dataset is corrupted by substituting it
with a training example of the same class drawn from the noisy dataset. The
amount of label noise can then be controlled by substituting a proportionate
number of training examples.

The clean dataset, ARCA23K-FSD, is a subset of FSD50K \cite{fsd50k_fonseca},
which is currently the largest clean dataset of sound events available. FSD50K
is comprised of more than \SI{40}{\kilo\relax} training examples and \num{200}
classes taken from the AudioSet ontology. In general, multiple labels are
associated with each audio clip.

For simplicity, we reduced FSD50K to a single-label dataset. First, clips
containing more than one type of sound were discarded. Next, to prevent class
overlap, classes that were ancestors of other classes (according to the
AudioSet ontology) were dropped, e.g. clips labelled as \textsf{Guitar} would
be dropped because \textsf{Acoustic guitar} and \textsf{Electric guitar} are
child classes. Finally, any sound class with an insufficient number of audio
clips was removed from the dataset. The thresholds are \num{50} instances in
the training set, \num{10} instances in the validation set, and \num{20}
instances in the test set. A total of \num{77} classes were retained after this
pruning process. Let \(\mathcal{L}\) denote the set of AudioSet labels that
remained.

\section{ARCA23K}
\label{sec:arca23k}

In this section, we describe how the clips in the ARCA23K dataset were retrieved
and curated. This dataset only includes a training set, since the validation set
and test set of the ARCA23K-FSD dataset are used for validation and testing,
respectively. We use a keyword-based algorithm to retrieve relevant clips and
label them accordingly. We will assume that we have access to the metadata of
every clip in the Freesound database and that we can download the clips. The
metadata includes a description of the clip and a set of \textit{tags} that are
intended to be search terms. As with FSD50K, we limit our search to clips that
are between \num{0.3} and \num{30} seconds \cite{fsd50k_fonseca}. After
curation, all clips are converted to 16-bit mono WAV files sampled at
\SI{44.1}{\kHz}.

\subsection{Retrieval Algorithm}
\label{ssec:retrieval}

The general framework for the retrieval algorithm is as follows. For every
candidate Freesound clip, the tags and description are tokenised and
preprocessed to give two word sequences, \(\dtags\) and \(\ddesc\), which we
refer to as \textit{documents}. For each label, \(l \in \mathcal{L}\), a
\textit{query}, \(q_l\) is constructed, which also involves tokenisation and
preprocessing. Given \(q_l\), \(\dtags\), and \(\ddesc\), a relevance score,
\(r(q_l,\dtags,\ddesc) \in [0,1]\), is computed, such that a higher score
indicates a better match between the query and the two documents. By computing
scores for each label, the most relevant label can be assigned to the clip:
\begin{equation}
  l^* \coloneqq \argmax_l r(q_l,\dtags,\ddesc).
\end{equation}
If \(r(q_{l^*},\dtags,\ddesc)\) is a low score, it indicates that none of the
labels are a good match according to the algorithm. For this reason, clips for
which \(r(q_{l^*},\dtags,\ddesc)<\tau\) are discarded, where \(\tau\) is a
predefined threshold.

\subsubsection{Tokenisation and Preprocessing}
\label{sssec:preprocess}

Tags, descriptions, and labels are tokenised and preprocessed using standard
practices in information retrieval \cite{ir_manning}. Tokenisation refers to
converting a sequence of characters into a sequence of words. During this
process, non-words such as punctuation and numbers are discarded. Tags are
already assumed to be a sequence of words, hence tokenisation is not necessary
for tags.

Preprocessing is carried out by first converting the words to lower-case so
that retrieval can be case-insensitive. Following this, lemmatisation is applied
to canonicalise words to their lemma form, e.g. `guitars' would be reduced to
`guitar'. In cases where the lemma depends on which word class the word belongs
to (e.g. verb, noun), the shortest lemma is chosen. For instance, `clapping'
would be reduced to `clap' because, even though `clapping' is the lemma for the
noun, `clap' is the lemma for the verb and is the shortest. After lemmatisation,
stop words such as conjunctions and prepositions are removed. Finally, any
duplicate words are also removed.

\subsubsection{Query Construction}
\label{sssec:query}

Given a label \(l\), a query, \(q_l\), is constructed as follows. The label is
first tokenised and preprocessed as per Section \ref{sssec:preprocess}. We will
refer to the resulting output as a \textit{root query} and denote it as
\(\bar{q}_l\). Next, a root query is constructed for every descendant label of
\(l\). For example, the label \textsf{Bowed string instrument} has several
descendants, such as \textsf{Cello} and \textsf{Double bass}. After
constructing the root queries, the final query \(q_l\) is constructed by
concatenating all of them.

\subsubsection{Computing Relevance Scores}
\label{sssec:rel_score}

In this section, we describe how relevance scores are computed. After creating a
query for each label \(l \in \mathcal{L}\), the vocabulary, \(V\), can be
defined as the concatenation of all the queries (after removing any duplicates).
After constructing \(V\), one can map any sequence of words, \(w\),
into a vector, \(v(w) \in \{0, 1\}^{|V|}\), such that
\begin{equation}
  \label{eq:vectorise}
  v(w)_i \coloneqq \left\{
    \begin{array}{ll}
      1 & w_i=V_i \\
      0 & \mathrm{otherwise} \\
    \end{array}
  \right.
\end{equation}
The relevance score is then defined as
\begin{equation}
  r(q,\dtags,\ddesc) \coloneqq \textrm{sim}(v(q), v(\dtags) + v(\ddesc)),
\end{equation}
where \(\textrm{sim}(a,b) \coloneqq (a \cdot b)/\norm{a}\norm{b}\) is the
cosine similarity between two vectors. The score is therefore a number between
\num{0} and \num{1}.

\subsubsection{Evaluation}
\label{sssec:evaluation}

The Freesound clips that are labelled by our retrieval algorithm include all
Freesound clips that are between \num{0.3} and \num{30} seconds in duration.
This means the clips that constitute the ARCA23K-FSD dataset are labelled by our
algorithm too. It is therefore possible to compare the labels assigned by our
algorithm to the ground truth labels of ARCA23K-FSD.

We used a threshold of \(\tau=0.5\), as it resulted in a reasonable compromise
between precision and recall. Our algorithm retrieved \SI{84.1}{\%} of the
ARCA23K-FSD clips and achieved an accuracy of \SI{90.3}{\%}. The accuracy was
greater than \SI{90}{\%} for \num{51} out of \num{77} classes, and the average
accuracy for these classes was found to be \SI{96.4}{\%}. For the other
\num{26} classes, the average accuracy was found to be \SI{67}{\%}.

It should be noted that the ARCA23K-FSD clips are not an unbiased sample of the
retrieved clips. The aim of the evaluation is not to determine label accuracy in
general but to demonstrate that it is comparable to approaches used for existing
datasets. In Section \ref{ssec:noise_rate}, we evaluate the accuracy of the
labels by manually verifying a subset of the ARCA23K dataset.

\subsection{Curation}
\label{ssec:curation}

After labelling the candidate Freesound clips using the retrieval algorithm, we
used a threshold of \(\tau=0.5\) to discard clips with a low relevance score.
All clips belonging to the FSD50K dataset were also discarded to prevent any
overlap. The number of retrieved clips at this point totalled almost
\SI{170}{\kilo\relax}. Next, the number of clips per class was reduced to
match ARCA23K-FSD, since our aim is to create a dataset that mirrors
ARCA23K-FSD. This was done by selecting a random sample of the correct size
from each class. For seven of the classes, there was an insufficient number of
clips to match the ARCA23K-FSD dataset, so the clips belonging to these classes
were dropped altogether. The same was done for ARCA23K-FSD, resulting in
\num{70} classes in total for both datasets.

\subsection{Noise Rate Estimation}
\label{ssec:noise_rate}

In this section, we describe how noise rates were estimated for the ARCA23K
dataset and present the results. The noise rate is defined as the percentage of
incorrectly labelled audio clips in the dataset. Similar to Fonseca et al.
\cite{fsd50k_fonseca}, we categorise clips as either `Present and predominant'
(PP), `Present but not predominant' (PNP), `Not present` (NP), and `Unsure'
(U). The reader is referred to the original work for detailed definitions
\cite{fsd50k_fonseca}. PNP and NP are further split based on whether the other
sounds are in-vocabulary (IV) or out-of-vocabulary (OOV). For example, NP/OOV
means that at least one OOV sound can be identified in the clip.

The noise rate of the dataset was estimated by selecting a random subset of the
dataset and performing listening tests. We selected \num{100} clips for the
sample and repeated the experiment three times with replacement. Each sample was
processed by a different listener, i.e. three listeners participated. The first
three authors of this paper carried out the tests. They were trained by
familiarising themselves with the classes, which involved reading the class
descriptions and listening to example clips. They were also able to listen to
example clips during the test and confer with each other\footnote{In practice,
listeners only conferred when they were unsure.}.

\begin{table}[t]
  \caption{Estimates of the proportion of ARCA23K clips that are PP/PNP/NP.
           The percentage of clips marked `Unsure' is \SI{1}{\%}.}
  \label{table:noise_rate}
  \centering
  \begin{tabularx}{0.45\textwidth}{lXXX}
    \toprule
        & PP         & PNP        & NP  \\
    \midrule
    IV  & \score{52.7}{5.8} & \score{2.3}{1.3} & \score{8.7}{3.5} \\
    OOV & N/A & \score{1.3}{0.7} & \score{33.3}{5.6} \\
    \bottomrule
  \end{tabularx}
\end{table}

The results are presented in Table \ref{table:noise_rate}. The noise rate can be
calculated by excluding the sounds categorised as U. When PNP sounds are
considered as incorrect, the noise rate was found to be \score{46.4}{4.8} (95\%
confidence interval). When PNP sounds are considered as correct, the noise rate
was found to be \score{42.4}{4.1}. Based on the results in Table
\ref{table:noise_rate}, \SI{75.9}{\%} of incorrectly labelled clips are OOV.
For many of the NP clips, we were able to identify them as NP from the tags and
description alone\footnote{All clips were listened to in their entirety
nonetheless.}, meaning that the labelling errors were the fault of the
retrieval algorithm; some were simple mistakes, while others required
understanding the context, which a keyword-based retrieval algorithm cannot
infer. In other cases, the uploaders' annotations were misleading or incorrect.
This was more prevalent with classes such as \textsf{Whoosh, swoosh, swish},
which are more open to interpretation without an agreed-upon definition.
Finally, we observed that many of the OOV sounds were quite similar in sound to
the IV classes. For example, \textsf{462351.wav}, labelled as \textsf{Acoustic
guitar}, contains sounds of a guitar string being strummed, but it is too
distorted to belong to any of the guitar classes.

\section{Experiments}
\label{sec:experiments}

In this section, we describe the experiments that were carried out and present
the results. Systems are evaluated using the accuracy and the mean average
precision (mAP). The mAP is approximately equal to the area under the
precision-recall curve; a higher value indicates better performance. We ran each
experiment five times and provide 95\% confidence intervals for the scores.

\subsection{System}
\label{sec:system}

The machine learning model used in our experiments is an 11-layer convolutional
neural network based on the VGG13 architecture \cite{vgg_simonyan}. Our model
differs from VGG13 in that it uses batch normalisation \cite{batchnorm_ioffe}
and only one fully-connected layer instead of three, as we found multiple
fully-connected layers to be unhelpful.

The model was trained with mel-spectrogram inputs. Prior to computing the
mel-spectrograms, the audio was downsampled from \SI{44.1}{\kHz} to
\SI{32}{\kHz}, which reduced the audio's data rate without significantly
affecting the results. The mel-spectrograms were then computed using a
\SI{32}{\ms} frame length, a \SI{16}{\ms} hop length, and \num{64} mel bins per
frame. Finally, the amplitudes of the mel-spectrograms were scaled
logarithmically.

Since the audio clips in both datasets vary in duration, we padded the clips
with silence. Instead of padding to a single fixed length, we used three
different lengths: \num{5} seconds, \num{15} seconds, and \num{30} seconds. The
least amount of padding was applied to each clip, e.g. a clip less than \num{5}
seconds would be padded to \num{5} seconds. When selecting clips for a
mini-batch, only clips of the same length were allowed. Without this
multi-length approach, each clip would have to be padded to the maximum length,
which would greatly increase training times.

The model was trained for \num{50} epochs using the cross-entropy loss function
and the AdamW optimiser \cite{adamw_loshchilov} with a learning rate of
\num{0.0005}, which was decayed by \SI{10}{\%} every two epochs. We used a batch
size of \num{64}, \num{32}, and \num{16} for \num{5}-, \num{15}-, and
\num{30}-second clips, respectively. By using different batch sizes, and given
the memory constraints, we were able to significantly improve training times and
even the classification accuracy. During inference, we averaged the predictions
of the top three epochs in order to reduce volatility.

\subsection{Adding Noise}
\label{ssec:add_noise}

In addition to training with the ARCA23K datasets, we also added synthetic label
noise to the ARCA23K-FSD dataset, which allows us to compare synthetic label
noise to the real-world label noise present in ARCA23K. The synthetic label
noise is closed-set rather than open-set. Let \(k \in \{1,\dots,K\}\) represent
the class associated with an instance, where \(K=70\) is the number of classes.
To add synthetic noise, we selected a proportion, \(\rho\), of training
examples and changed the class \(k\) of each selected example to
\begin{equation}
  (k + i) \mod K,
\end{equation}
where \(i\) is a random integer drawn from a suitable distribution. Two types
of label noise were considered: uniform and class-conditional. In the case of
uniform noise, \(i\) followed the uniform distribution, \(U(1,K-1)\). In the
case of class-conditional noise, the geometric distribution was used with
\(p=0.5\). This distribution is concentrated over a small number of outcomes,
which is realistic because only a small set of classes tend to be incorrectly
attributed to a sound.

Finally, we ran experiments in which we replaced a proportion, \(\rho\), of the
ARCA23K-FSD training examples with ARCA23K training examples. Recall from
Section \ref{sec:arca23k-fsd} that this is equivalent to controlling the noise
rate of ARCA23K. For each example that was replaced, the replacement example was
restricted to be identically labelled. The noise rate of the resulting mixed
dataset is a fraction of the noise rate estimated in Section
\ref{ssec:noise_rate}. For example, \(\rho=0\) corresponds to a noise rate of
\num{0}, \(\rho=1\) corresponds to a noise rate of \SI{46.4}{\%}, and
\(\rho=0.5\) corresponds to a noise rate of \SI{23.2}{\%}.

\subsection{Representation Learning}
\label{ssec:repr}

As a final set of experiments, we examined how label noise affects
the representations that are learned. To do this, we trained a linear classifier
on embeddings derived from the output of the VGG model's penultimate layer and
evaluated its performance. The VGG model was first trained as normal using a
noisy dataset (either ARCA23K or ARCA23K-FSD with synthetic label noise). Next,
using the output of the penultimate layer as input data, a linear classifier was
trained on the (clean) ARCA23K-FSD dataset. In addition to training with the
whole of ARCA23K-FSD, we trained the linear classifier with \SI{10}{\%},
\SI{20}{\%}, and \SI{50}{\%} of the dataset.

\subsection{Results}

\begin{table}[t]
  \caption{Model performance when using different training sets.}
  \label{table:main_results}
  \centering
  \begin{tabularx}{0.45\textwidth}{lXX}
    \toprule
    Dataset            & Accuracy  & mAP \\
    \midrule
    ARCA23K            & \score{50.08}{0.78}    & \score{52.32}{0.77} \\
    ARCA23K-FSD        & \score{61.16}{0.41}    & \score{66.28}{0.59} \\
    Uniform Noise      & \score{38.12}{1.83}    & \score{35.76}{2.10} \\
    Conditional Noise  & \score{38.35}{0.56}    & \score{36.82}{0.62} \\
    \bottomrule
  \end{tabularx}
\end{table}

The first group of results are presented in Table \ref{table:main_results}. In
this table, we compare the performance of the system when trained using: (1)
ARCA23K, (2) ARCA23K-FSD, (3) ARCA23K-FSD but with uniform label noise, and (4)
ARCA23K-FSD but with conditional label noise. We set \(\rho=0.45\) for both (3)
and (4). The results show that training with ARCA23K-FSD gives an mAP score of
\SI{66.28}{\%}, which is \num{14}{\%} higher than when training with ARCA23K,
which suggests that the presence of open-set label noise has a considerable
effect on training. However, it can be seen that the effect of synthetic label
noise is much more severe, as the mAP drops below \num{40}{\%}.

We hypothesise that there are at least two reasons why real-world, open-set
label noise has a milder effect on training. First, we believe that OOV clips
are inherently less likely to harm performance compared to mislabelled IV clips,
especially if the OOV clips sound very different to the IV clips. Recall that
most of the incorrectly labelled clips in ARCA23K are OOV. Second, as observed
in Section \ref{ssec:noise_rate}, a considerable number of OOV clips were found
to be similar in sound to the IV clips. If these clips are labelled accordingly
(e.g. \textsf{462351.wav} labelled as \textsf{Acoustic guitar}), they can be
considered as surrogates of the IV clips. Rather than being detrimental to
learning, these surrogates are likely to be beneficial. Both of these
hypotheses were verified to some extent in previous work \cite{ood_iqbal}.

\begin{figure}[!t]
  \centering
  \includegraphics[width=0.47\textwidth]{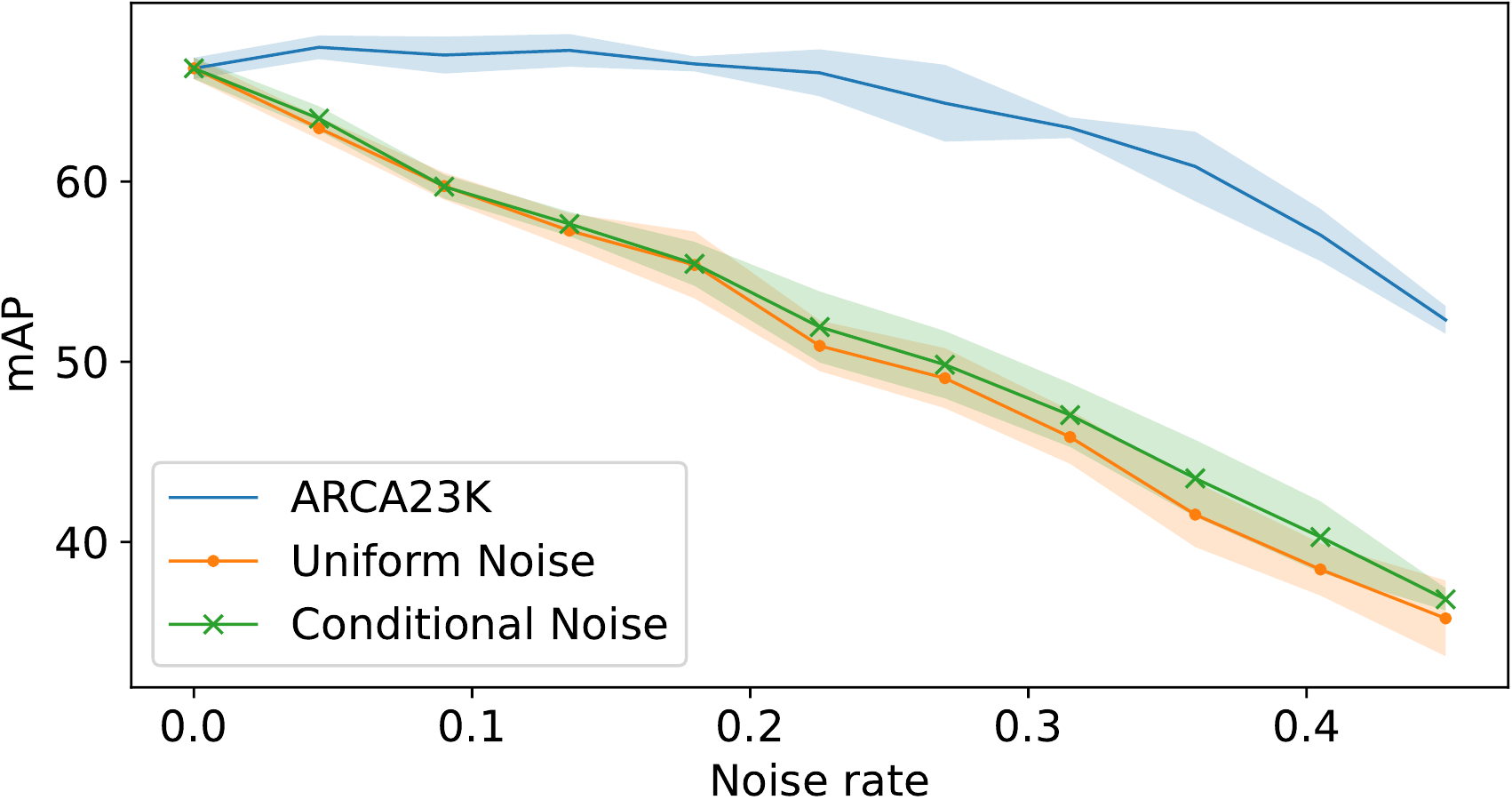}
  \caption{The mAP scores as \(\rho\) is varied from \num{0} to \num{0.45}.}
  \label{fig:results}
\end{figure}

In Figure \ref{fig:results}, we present the mAP scores as \(\rho\) (refer to
Section \ref{ssec:add_noise}) is varied from \num{0} to \num{0.45} at increments
of \num{0.045}. For all three types of noise, the performance generally
decreases as \(\rho\) increases. However, while the plots appear linear for
synthetic label noise, the plot for real-world, open-set label noise is
non-linear. The performance decreases exponentially as the noise rate increases,
albeit it is roughly the same until the noise rate exceeds \SI{20}{\%}.

\begin{table}[t]
  \caption{mAP scores for the linear classifier. Columns indicate the percentage
           of ARCA23K-FSD clips used for training.}
  \label{table:repr_results}
  \centering
  \begin{tabularx}{0.45\textwidth}{lXXXX}
    \toprule
    Dataset           & \SI{10}{\%}    & \SI{20}{\%}    & \SI{50}{\%}    & \SI{100}{\%} \\
    \midrule
    ARCA23K           & \SI{55.27}{\%} & \SI{58.94}{\%} & \SI{58.91}{\%} & \SI{59.82}{\%} \\
    Uniform Noise     & \SI{30.86}{\%} & \SI{37.11}{\%} & \SI{42.29}{\%} & \SI{45.52}{\%} \\
    Conditional Noise & \SI{41.15}{\%} & \SI{46.06}{\%} & \SI{48.11}{\%} & \SI{50.09}{\%} \\
    Random Weights    & \SI{7.93}{\%}  & \SI{10.11}{\%} & \SI{13.12}{\%} & \SI{16.23}{\%} \\
    \bottomrule
  \end{tabularx}
\end{table}

The results for the experiments described in Section \ref{ssec:repr} are
presented in Table \ref{table:repr_results}. We have also reported the
performance when using a randomly-initialised VGG model to compute the
embeddings. Similar to the results in Table \ref{table:main_results}, the
performance is considerably worse when using synthetic label noise. When using
ARCA23K to learn the representation, the performance of the linear classifier is
relatively high even when training with \SI{10}{\%} of ARCA23K-FSD. On the other
hand, the scores are still significantly lower than the score of \SI{66.28}{\%}
in Table \ref{table:main_results}. These results show that label noise has a
substantial effect on the quality of the learned representations.

\section{Conclusion}

In this paper, we introduced the ARCA23K dataset along with the companion
ARCA23K-FSD dataset, which were created with the intention of studying open-set
label noise. ARCA23K was created with minimal human labour by retrieving and
curating clips from the Freesound database using an automated process, while
ARCA23K-FSD was derived from FSD50K. We described the dataset creation process
in detail and characterised the type of label noise present in ARCA23K via
listening tests. Using these datasets, we were able to study the effect of
label on learning in a controlled manner. We found that, while open-set label
noise negatively affected performance, the impact was considerably milder than
that of synthetic label noise. Furthermore, our experiments showed the extent to
which label noise affects the learned representations of a model.

\section{Acknowledgement}

This work was funded by the Engineering and Physical Sciences Research Council
(EPSRC) Doctoral Training Partnership grants EP/N509772/1 and EP/R513350/1. It
was also supported in part by EPSRC project EP/T019751/1 and by a Newton
Institutional Links Award from the British Council with grant number 623805725.

\bibliographystyle{IEEEtran}
\bibliography{references}

\begin{thebibliography}{10}
\providecommand{\url}[1]{#1}
\def\UrlFont{\rmfamily}
\providecommand{\newblock}{\relax}
\providecommand{\bibinfo}[2]{#2}
\providecommand\BIBentrySTDinterwordspacing{\spaceskip=0pt\relax}
\providecommand\BIBentryALTinterwordstretchfactor{4}
\providecommand\BIBentryALTinterwordspacing{\spaceskip=\fontdimen2\font plus
\BIBentryALTinterwordstretchfactor\fontdimen3\font minus
  \fontdimen4\font\relax}
\providecommand\BIBforeignlanguage[2]{{%
\expandafter\ifx\csname l@#1\endcsname\relax
\typeout{** WARNING: IEEEtran.bst: No hyphenation pattern has been}%
\typeout{** loaded for the language `#1'. Using the pattern for}%
\typeout{** the default language instead.}%
\else
\language=\csname l@#1\endcsname
\fi
#2}}

\bibitem{freesound_fonseca}
E.~Fonseca, J.~Pons, X.~Favory, F.~Font, D.~Bogdanov, A.~Ferraro, S.~Oramas,
  A.~Porter, and X.~Serra, ``Freesound datasets: A platform for the creation of
  open audio datasets,'' in \emph{Proc. 18th Int. Society Music Information
  Retrieval Conf. (ISMIR)}, Suzhou, China, 2017, pp. 486--493.

\bibitem{audioset_gemmeke}
J.~F. Gemmeke, D.~P.~W. Ellis, D.~Freedman, A.~Jansen, W.~Lawrence, R.~C.
  Moore, M.~Plakal, and M.~Ritter, ``{Audio Set}: An ontology and human-labeled
  dataset for audio events,'' in \emph{2017 IEEE International Conference on
  Acoustics, Speech and Signal Processing (ICASSP)}, New Orleans, LA, USA,
  2017, pp. 776--780.

\bibitem{esc_piczak}
K.~J. Piczak, ``{ESC}: Dataset for environmental sound classification,'' in
  \emph{Proceedings of the 23rd ACM International Conference on Multimedia},
  New York, NY, USA, 2015, pp. 1015--1018.

\bibitem{urbansound_salamon}
J.~Salamon, C.~Jacoby, and J.~P. Bello, ``A dataset and taxonomy for urban
  sound research,'' in \emph{Proceedings of the 22nd {ACM} International
  Conference on Multimedia}, Orlando, FL, USA, 2014, pp. 1041--1044.

\bibitem{fsd50k_fonseca}
E.~{Fonseca}, X.~{Favory}, J.~{Pons}, F.~{Font}, and X.~{Serra}, ``{FSD50K}: An
  open dataset of human-labeled sound events,'' \emph{arXiv preprint
  arXiv:2010.00475}, Oct{.} 2020.

\bibitem{noise_zhu}
B.~Zhu, K.~Xu, Q.~Kong, H.~Wang, and Y.~Peng, ``Audio tagging by cross
  filtering noisy labels,'' \emph{IEEE/ACM Transactions on Audio, Speech, and
  Language Processing}, vol.~28, pp. 2073--2083, Jul{.} 2020.

\bibitem{vggsound_chen}
H.~Chen, W.~Xie, A.~Vedaldi, and A.~Zisserman, ``{VGGSound}: A large-scale
  audio-visual dataset,'' in \emph{IEEE International Conference on Acoustics,
  Speech and Signal Processing (ICASSP)}, 2020, pp. 721--725.

\bibitem{fsdkaggle2018_fonseca}
E.~Fonseca, M.~Plakal, F.~Font, D.~P. Ellis, X.~Favory, J.~Pons, and X.~Serra,
  ``General-purpose tagging of {Freesound} audio with {AudioSet} labels: Task
  description, dataset, and baseline,'' in \emph{Proceedings of the Detection
  and Classification of Acoustic Scenes and Events 2018 Workshop (DCASE2018)},
  Woking, UK, 2018, pp. 69--73.

\bibitem{fsdkaggle2019_fonseca}
E.~Fonseca, M.~Plakal, F.~Font, D.~P. Ellis, and X.~Serra, ``Audio tagging with
  noisy labels and minimal supervision,'' in \emph{Proceedings of the Detection
  and Classification of Acoustic Scenes and Events 2019 Workshop (DCASE2019)},
  New York, NY, USA, 2019, pp. 69--73.

\bibitem{fsdnoisy18k_fonseca}
E.~{Fonseca}, M.~{Plakal}, D.~P.~W. {Ellis}, F.~{Font}, X.~{Favory}, and
  X.~{Serra}, ``Learning sound event classifiers from web audio with noisy
  labels,'' in \emph{International Conference on Acoustics, Speech and Signal
  Processing (ICASSP)}, Brighton, UK, 2019, pp. 21--25.

\bibitem{noise_patrini}
G.~Patrini, A.~Rozza, A.~Menon, R.~Nock, and L.~Qu, ``Making deep neural
  networks robust to label noise: A loss correction approach,'' in
  \emph{IEEE/CVF Conference on Computer Vision and Pattern Recognition (CVPR)},
  Honolulu, HI, USA, 2017, pp. 2233--2241.

\bibitem{noise_sukhbaatar}
S.~Sukhbaatar, J.~Bruna, M.~Paluri, L.~Bourdev, and R.~Fergus, ``Training
  convolutional networks with noisy labels,'' in \emph{International Conference
  on Learning Representations}, San Diego, CA, USA, 2015.

\bibitem{ood_iqbal}
T.~Iqbal, Y.~Cao, Q.~Kong, M.~D. Plumbley, and W.~Wang, ``Learning with
  out-of-distribution data for audio classification,'' in \emph{ICASSP 2020 -
  2020 IEEE International Conference on Acoustics, Speech and Signal Processing
  (ICASSP)}, 2020, pp. 636--640.

\bibitem{noise_kumar}
A.~Kumar, A.~Shah, A.~Hauptmann, and B.~Raj, ``Learning sound events from webly
  labeled data,'' in \emph{Proceedings of the 28th International Joint
  Conference on Artificial Intelligence (IJCAI)}, Macao, China, 2019, pp.
  2772--2778.

\bibitem{noise_jiang}
L.~Jiang, D.~Huang, M.~Liu, and W.~Yang, ``Beyond synthetic noise: Deep
  learning on controlled noisy labels,'' in \emph{Proceedings of the 37th
  International Conference on Machine Learning (ICML)}, vol. 119, 2020, pp.
  4804--4815.

\bibitem{ir_manning}
C.~D. Manning, P.~Raghavan, and H.~Schütze, \emph{Introduction to information
  retrieval}.\hskip 1em plus 0.5em minus 0.4em\relax Cambridge University
  Press, 2008.

\bibitem{vgg_simonyan}
K.~Simonyan and A.~Zisserman, ``Very deep convolutional networks for
  large-scale image recognition,'' in \emph{International Conference on
  Learning Representations (ICLR)}, San Diego, CA, USA, 2015.

\bibitem{batchnorm_ioffe}
S.~Ioffe and C.~Szegedy, ``Batch normalization: Accelerating deep network
  training by reducing internal covariate shift,'' in \emph{Proceedings of the
  32nd International Conference on Machine Learning (ICML)}, vol.~37, Lille,
  France, 2015, pp. 448--456.

\bibitem{adamw_loshchilov}
I.~Loshchilov and F.~Hutter, ``Decoupled weight decay regularization,'' in
  \emph{International Conference on Learning Representations (ICLR)}, New
  Orleans, LA, USA, 2019.

\end{thebibliography}

\end{sloppy}

\end{document}